\documentclass[prl, twocolumn]{revtex4}
\usepackage{amsfonts}
\usepackage{amsmath}
\usepackage{amssymb}
\usepackage{graphicx}
\usepackage{epstopdf}
\usepackage{color}
\input epsf.sty 

\begin{document}

\title{Emergence of fermionic finite-temperature critical point in a Kondo lattice}
\author{Po-Hao Chou$^{1}$, Liang-Jun Zhai$^{1}$, Chung-Hou Chung$^{2}$, Chung-Yu Mou$^{1,3,4}$, and Ting-Kuo Lee$^{3}$ 
}
\affiliation{$^{1}$Department of Physics, National Tsing Hua University, Hsinchu 30043,
Taiwan, 300, R.O.C.}
\affiliation{$^{2}$Electrophysics Department, National Chiao-Tung University, HsinChu, Taiwan, R.O.C.}
\affiliation{$^{3}$Institute of Physics, Academia Sinica, Nankang, Taiwan, R.O.C.}
\affiliation{$^{4}$Physics Division, National Center for Theoretical Sciences, P.O.Box
2-131, Hsinchu, Taiwan, R.O.C.}

\begin{abstract}
The underlying Dirac point is central to the profound physics manifested in a wide class of materials. However, it is often difficult to drive a system with Dirac points across the massless fermionic critical point. Here by exploiting screening of local moments under spin-orbit interactions in a Kondo lattice, we show that below the Kondo temperature, the Kondo lattice undergoes a topological transition from a strong topological insulator to a weak topological insulator at a finite temperature $T_D$. At $T_D$, massless Dirac points emerge and the Kondo lattice becomes a Dirac semimetal. Our analysis indicates that the emergent relativistic symmetry dictates non-trivial thermal responses over large parameter and temperature regimes. In particular, it yields critical scaling behaviors both in magnetic and transport responses near $T_D$.
\end{abstract}

\maketitle

Since the discovery of graphene\cite{Geim1,Geim2}, enormous efforts were inspired to search materials with similar Dirac-like electronic structures. It was soon realized that a variety of materials, known as topological insulators, can exhibit low-dimensional Dirac Fermions at surfaces\cite{Kane, Zhang, silicene} due to nontrivial topology in bulk electronic structures. More recently, 3D Dirac semi-metallic phases are also found in Na$_3$Bi\cite{Na3Bi} and Cd$_3$As$_2$\cite{Cd2As3_1,Cd2As3_2}. In these materials, the underlying Dirac points are central to the associated novel properties\cite{Wehling} and the mass of the Dirac Fermions is the energy gap that controls the transition between the topological trivial and the topological nontrivial phases\cite{Shen, Zhang, Sato}.  Right at the point when the mass vanishes, the material is a Dirac semimetal which is at the fermionic quantum critical point (QCP) between the hole Fermi liquid and the electron Fermi liquid\cite{Sheehy}. The quantum criticality extends effects of the Dirac point to a finite critical regime\cite{Sheehy} and results in nontrivial scalings in Dirac semimetals. Despite the profound physics that can be manifested in Dirac semimetals, the access of the critical point requires particular symmetries \cite{Kane2012} and they are rare in real materials. Furthermore, the transition across the critical point requires tunability of electronic structures. Successful manipulations of electronic states across the critical point are often difficult\cite{Sato} and are usually not performed in the same system. It is therefore desirable to search for feasible ways to access the fermionic critical point.

In this work, we explore topological phases at finite temperatures due to the many-body screening interaction of localized spins and conduction electrons in a Kondo lattice. We demonstrate that the hybridization of localized spins and conduction electrons leads to temperature-dependent electronic energy bands with the mass of the Dirac fermions being tunable. When spin-orbit interactions are included in hybridization, we find that the Kondo lattice is a strong topological insulator at low temperature\cite{Coleman} and undergoes a topological transition to a weak topological insulator at a higher temperature $T_ D$. At $T_D$, Dirac points emerge and the system is a Dirac semimetal. Our results indicate that the finite temperature  transition through a Dirac semi-metallic phase results in nontrivial critical scaling behaviors both in transport and magnetic responses near $T_D$.

{\it The model --} We start with the Anderson lattice Hamiltonian (ALH) on a cubic lattice, which is shown to characterize the topological Kondo insulating phase of SmB$_6$\cite{Coleman} 
\begin{eqnarray} \label{Eq1}
H &=& \sum_{\mathbf{k} \sigma} (\xi_{\mathbf{k}} c_{\mathbf{k}\sigma}^\dag c_{\mathbf{k} \sigma} + \xi^d_{\mathbf{k}} d_{\mathbf{k} \sigma}^\dag d_{\mathbf{k} \sigma}) \nonumber \\ 
&+& \sum_{\mathbf{k} \sigma \sigma'}  (V^{\sigma \sigma'}_{\mathbf k} c_{\mathbf{k} \sigma}^\dag d_{\mathbf{k} \sigma'}+ H.C.)  + U\sum_i n_{i\uparrow}^dn_{i\downarrow}^d. \label{ALH}
\end{eqnarray}
Here $c^\dag$ and $d^\dag$ creates conduction and more localized electrons in f orbit respectively. $\xi_{\mathbf k}$ is equal to $\varepsilon_{\mathbf k}-\mu$ with $\varepsilon_{\mathbf k} =-2t\sum_{i=x,y,z} cosk_i-4t'\sum_{i\neq j} cosk_icosk_j$ and $\mu$ being the chemical potential. $\xi^d_{\mathbf{k}} = \varepsilon_d - \eta \varepsilon_{\mathbf k} -\mu$ characterizes the narrow band formed by $d$ electrons with $\eta$ being the bandwidth and $\varepsilon^d$ being the relative shift of band center. $\mathbf{V}_{\mathbf{k}}$ is the hybridization between $c$ and $d$ electrons, given by  $v_0 I+2\lambda_{so} \sum_{i=x,y,z} \sigma_i \sin k_i$ with $I$ and $\sigma_i$ being the unit and the Pauli matrices respectively. Here $v_0$ vanishes due to odd parity of the $f$ orbits\cite{Coleman} so that the spin-orbit interaction $\lambda_{so}$ dominates. Finally, $U$ describes the Hubbard repulsion between $d$ electrons. 

{\it Kondo screening.--} We first analyze the screening interaction between $c$ and $d$ electrons. In the presence of $\lambda_{so}$, effective spin interactions are modified.  By applying the Schrieffer-Wolff transformation\cite{SW} on ALH in the continuum limit with $\varepsilon_d-\mu=-U/2$\cite{sup}, we
obtain the generalized Kondo lattice Hamiltonian\cite{KLH} with three type of spin interactions represented by (i) Heisenberg interaction $J_1\vec{s}_{\mathbf{k} \mathbf{k}'}\cdot\vec{s}_d$, (ii) Dzyaloshinskii-Moriya interaction $J_2[-i(\vec{k}-\vec{k'})\cdot(\vec{s}_{kk'}\times\vec{s}_d)]$, and (iii) tensor type interaction $J_3[ (\vec{s}_{kk'}\cdot\vec{k})(\vec{k'}\cdot\vec{s}_d)+ (\vec{s}_{kk'}\cdot\vec{k'})(\vec{k}\cdot\vec{s}_d)]$, where $\vec{s}_{\mathbf{k} \mathbf{k}'}=\psi_{\mathbf{k}}^\dag\vec{\sigma}\psi_{\mathbf{k}'}$ and $\vec{s}_{d}=\phi_d^\dag\frac{\vec{\sigma}}{2} \phi_{d}$  with $\psi_{\mathbf{k}} = (c_{\mathbf{k} \uparrow}, c_{\mathbf{k} \downarrow})$ and $\phi_{d}^\dag= (d_{\uparrow}^\dag, d_{\downarrow}^\dag)$.  The strengths of spin interactions are given by
$J_1= J v_0^2$, $J_2=J v_0\lambda_{so}$, $J_3=4 J \lambda_{so}^2$ with  $J =1/(U+\varepsilon_d-\mu)-1/(\varepsilon_d-\mu)$. The screening of a single localized spin is analyzed by a renormalization group (RG) analysis\cite{poorman}. By defining dimensionless parameters $g_i=2\rho J_i$ with $\rho$ being the density of states at the Fermi energy, we obtain coupled flow equations\cite{sup}:
$\dot{g_1}=-(g_1^2+g_2^2)$, $\dot{g_2}=-g_2(g_1+g_3)$,
$\dot{g_3}=-(g_3^2+g_2^2)$.
Here the solution is: $(g_1-g_3)/(g_2)= $ const. The initial coupling constants at the band cutoff $D_0$ satisfy $g_1(D_0)g_3(D_0)=g_2^2(D_0)$. We find that all the $g_i$'s flow to infinity at the Kondo temperature\cite{sup}
%$\dot{u}=u^2$
%$\dot{v}=v^2$ with
%$u=\frac{g_1+g_3}{2}+\sqrt{(\frac{g_1-g_3}{2})^2+g_2^2}$ and $v=\frac{g_1+g_3}{2}-\sqrt{(\frac{g_1-g_3}{2})^2+g_2^2}$. 
\begin{equation}
T_K =D_0 e^{\frac{-1}{2\rho(J_1+J_3)}}. \label{T_K}
\end{equation}
$T_K$ is clearly enhanced in comparison to the Kondo temperature without $\lambda_{so}$, $T^0_K =D_0 e^{\frac{-1}{2\rho J_1}}$. This enhancement persists even when many localized spins are included\cite{sup}.

$\lambda_{so}$ also changes the screening scenario below $T_K$. This is exhibited by performing the decomposition: $C_{\mathbf{k}\sigma}=\frac{1}{k}\sum_{lm}\mathrm{Y}^m_l(\hat{k})C^m_{lk\sigma}$ with $Y^m_l$ being spherical harmonics. Because the hybridization only mixes orbital angular momentum $l=1$ with spins, the $d_{\uparrow /\downarrow}$ electron only couples $c$ electrons with $j_z = \pm 1/2$ through a particular linearly-combined $C^{m \dag}_{1k\sigma}$ defined by $ a^\dag_{k,j_z=\pm 1/2} = v_0C_{0 k,\pm 1/2}^{0\dag}\pm v_1( C_{1 k,\pm\frac{1}{2}}^{0\dag} - \sqrt{2}C_{1 k,\mp\frac{1}{2}}^{\pm1\dag})$ with $v_1=\frac{2}{\sqrt{3}}\lambda_{so}k_F$\cite{sup}. The resulting spin interaction is $J_1 \vec{s}_{\mathbf{k} \mathbf{k}'} \cdot \vec{s}_d$ with $\vec{s}_{\mathbf{k} \mathbf{k}'}= \sum_{\alpha,\beta} a^{\dag}_{k, \alpha} \vec{\sigma}_{\alpha \beta} a_{k,\beta}$. Hence it is the total angular momenta of  $c$ electrons that interact with the $d$ electron. Below $T_K$, $c$ and $d$ electrons are coupled with the total angular momentum being screened.  However, since $d$ electrons only couples to $ a^\dag_{k,j_z=\pm 1/2}$, other linearly combinations of $C^{m \dag}_{1k\sigma}$ are left free\cite{sup}. It hints that there may exist more structures in the phase space below $T_K$.

\begin{figure}[th]
\includegraphics[height=1.4in,width=2.3in] {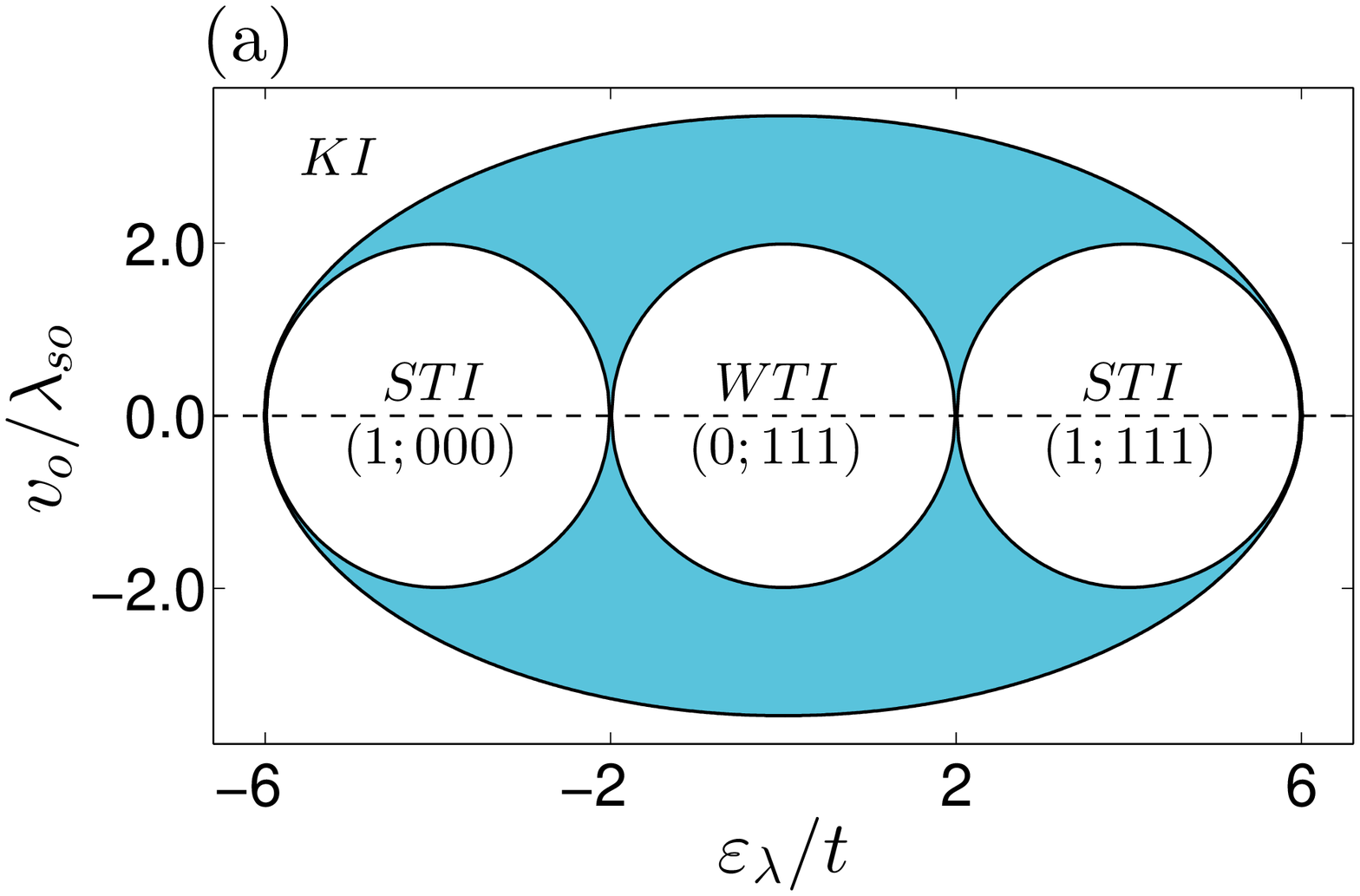}
\includegraphics[height=1.4in,width=2.3in] {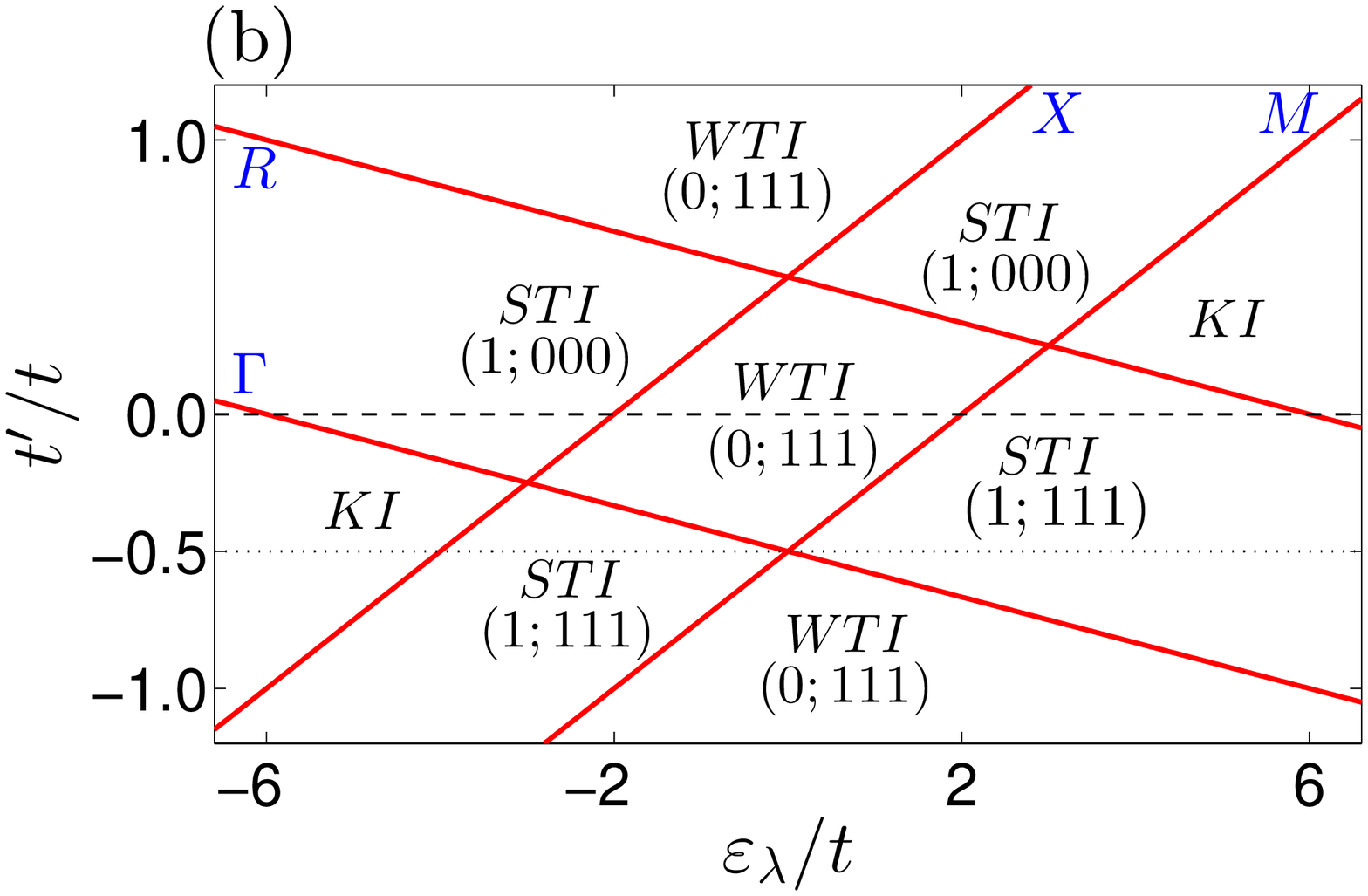}
\caption{(a) Topological phase diagram of the Kondo lattice with $t'=0$. Here $\varepsilon_{\lambda}  \equiv \frac{\varepsilon_d+\lambda}{1+\eta r^2}$ and $(\nu_0;\nu_1, \nu_2,\nu_3)$ are topological indices. Shaded regimes are gapless phases and white regimes are phases with gaps in electronic structures, labelled by strong topological insulator (STI), weak topological insulator (WTI), and Kondo insulator (KI) when the valence bands are filled. The gapless phases at $\varepsilon_{\lambda}/t=-6,-2,2,6$ are Dirac semi-metallic phases with corresponding Dirac points being at time reversal momenta $\Gamma, X, M, R$ respectively. (b) Topological phases for $t' \neq 0$ and $v_0=0$. Here solid lines, labelled by $\Gamma$, $X$, $M$, and $R$, are phase boundaries with energy gap vanishing at $\Gamma$, $X$, $M$, and $R$ respectively.}
\label{fig1}
\end{figure}
\begin{figure}[t]
%\includegraphics[width=1.1\linewidth]{Fig2.pdf}
\includegraphics[height=1.7in,width=2.2in] {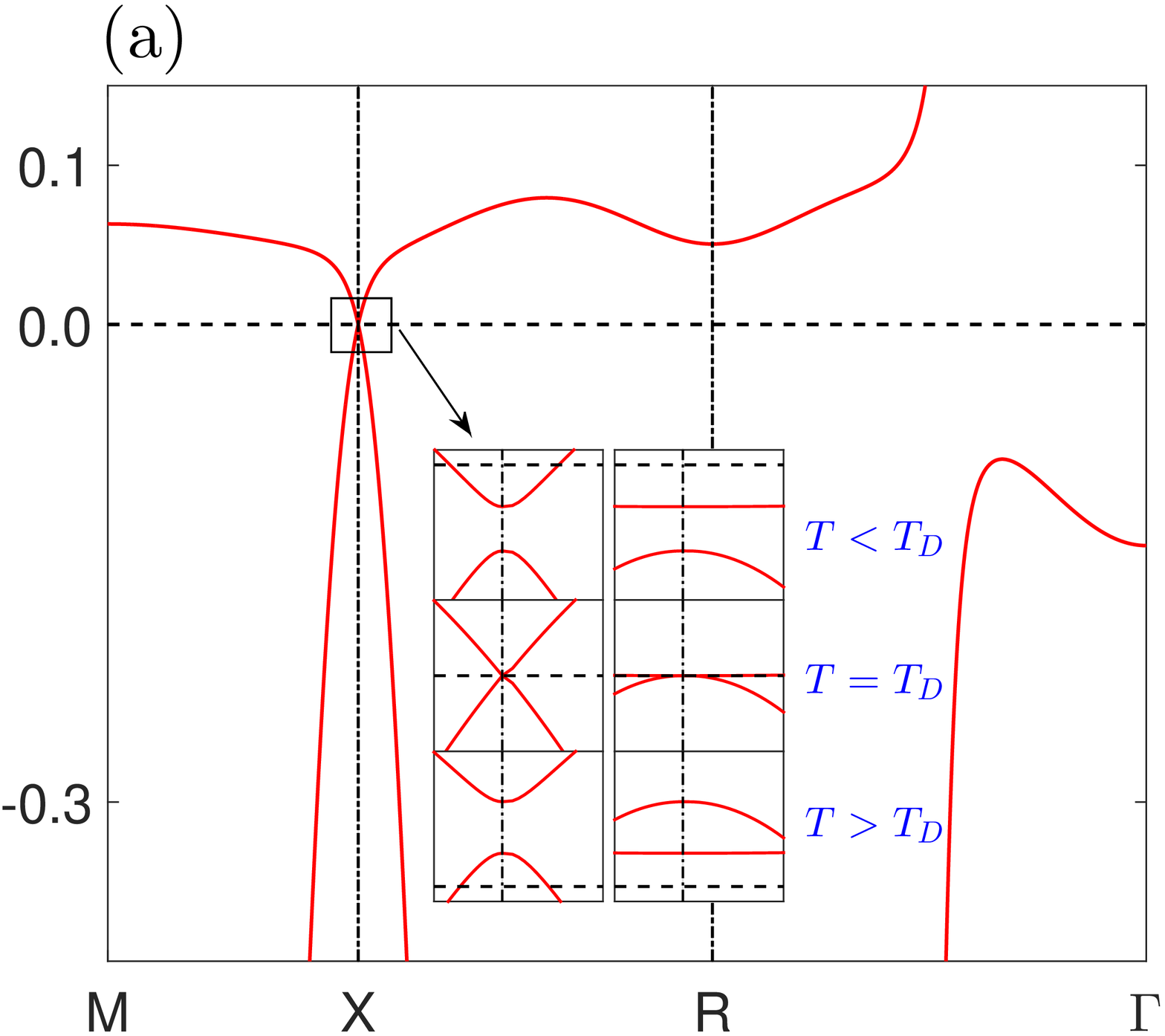}
\includegraphics[height=2.0in,width=2.3in] {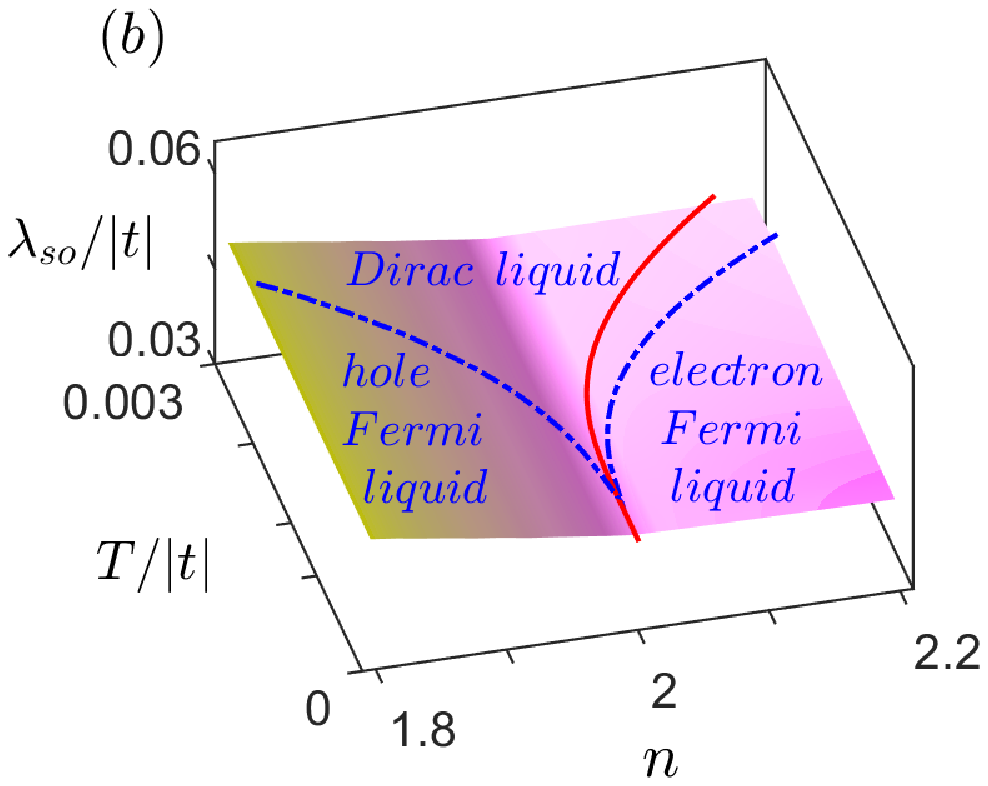}
\caption{(a) Emergence of a finite-temperature Dirac point at $X$ point with $n=2.074, t=-1,t'=-0.3, \lambda_{so}=0.2, \eta=0.05, \varepsilon_d/t=0.449$. Inset shows relative band positions before hybridization, indicating the occurrence of a band-inversion at $T_D$. (b) Typical surface of $T^*_D$ as a function of $n$ and $\lambda_{so}$ with $t=-1,t'=0.3, \eta=0.001, \varepsilon_d/t=3.19$. The red line marks critical temperatures $T_D$ at which the chemical potential is at the Dirac point. The blue dash lines indicate the crossover temperature $T^*$ that separates the Dirac liquid regime from the Fermi liquid regime.  As a reference, by using Eq.(\ref{T_K}), $T_K$ can be estimated by $4t \exp [-9\eta t^2/(2\lambda^2_{so})]$, which are in the range $0.026t-0.24t$ for $\lambda_{so}/t=0.03-0.04$ and are at least the order of $10T_D$.
}\label{TD}
\end{figure}
%\begin{figure}[ht]
% \includegraphics[height=1.6in,width=2.5in] {Fig3.eps}
%\caption{Typical surface of the temperature, $T^*_D$, for the occurrence of Dirac points as a function of $n$ and $\lambda_{SO}$. The red line marks critical temperatures $T_D$ at which the chemical potential is at the Dirac point. The blue dash lines indicate the crossover temperature $T^*$ that separates the Dirac liquid regime from the Fermi liquid regime. %(c) Projection of  $T^*_D$ on $T$ and $\lambda_{SO}$ plane. Here solid lines are contours of %$T^*_D$ with $n$ increasing from 1.96 to 2.08 and the red dash line is the projection of $T_D$. %Parameters: $t=-1, t'=-0.15, \varepsilon_d/t=1.37, \eta=0.02$, and $\mu \sim 1.395-1.405$. 
%}\label{phase}
%\end{figure}
{\it Topological phase diagram.--} To access electronic structures below $T_K$ in
the large $U$ limit, we apply the slave-boson method by expressing $d_{i\sigma}^{\dag}=f_{i\sigma}^{\dag}b_i$, where $f_i$ and $b_i$ are spinon and holon operators satisfying the constraint, $\sum_{\sigma}f_{i\sigma}^{\dag}f_{i\sigma}+b_i^{\dag}b_i = 1$. The constraint is removed by a Lagrangian field $\lambda_i$ so that the Hamiltonian has to include the extra term $\sum_i\lambda_i(\sum_{\sigma}f_{i\sigma}^{\dag}f_{i\sigma}+b_i^{\dag}b_i -1)$. In the mean field approximation, holons condense with $<b_i>=<b_i^{\dag}> \equiv r$ and $\lambda_i$ is replaced its mean-field value $\lambda$. The Hamiltonian becomes
$H_M =\sum_{\mathbf{k} \sigma}  \left( c_{\mathbf{k} \sigma}, f_{\mathbf{k} \sigma} \right)^\dag H_{\mathbf{k}} \left(  c_{\mathbf{k} \sigma}, f_{\mathbf{k} \sigma} \right) + N \lambda (r^2-1)$ with
\begin{equation}\label{2}
H_{\mathbf{k}}= \begin{pmatrix} \xi_{\mathbf{k}} I& r \mathbf{V}_{\mathbf{k}}  \\
           r \mathbf{V}_{\mathbf{k}}  & \tilde{\xi}^d_{\mathbf{k}}I
                       \end{pmatrix}.
 \end{equation}
Here $N$ is number of sites and $\tilde{\xi}^d_{\mathbf{k}}=(\varepsilon_d+\lambda)-\eta r^2\varepsilon_{\mathbf{k}}-\mu$. It is convenient to rewrite $\xi_{\mathbf{k}} = m_{\mathbf{k}} - \mu_{\mathbf{k}}$ and $\tilde{\xi}^d_{\mathbf{k}}= -m_{\mathbf{k}} - \mu_{\mathbf{k}}$ with $m_{\mathbf{k}}=\frac{(1+\eta r^2 )\varepsilon_{\mathbf{k}}-\varepsilon_d-\lambda}{2}$ and  $\mu_{\mathbf{k}}=\mu-\frac{(1-\eta r^2)\varepsilon_{\mathbf{k}}+\varepsilon_d+\lambda}{2}$.
By minimizing the free energy, $r$ and $\lambda$ are determined self-consistently through the
mean-field equations
\begin{eqnarray}
& & \frac{1}{N} \sum_{\mathbf{k} \sigma}  \langle f_{\mathbf{k} \sigma}^{\dag}   f_{\mathbf{k} \sigma}  \rangle
+r^2=1, \label{mean1} \\
& & \frac{1}{N} \sum_{\mathbf{k} \sigma \sigma'}   \left[ {\rm Re}  (V^{\sigma \sigma'}_{\mathbf {k}} \langle c_{\mathbf{k} \sigma}^{\dag}  f_{\mathbf{k} \sigma'} \rangle ) -r \eta \varepsilon_{\mathbf{k}} \delta _{\sigma\sigma'}  \langle f_{\mathbf{k} \sigma}^{\dag} f_{\mathbf{k} \sigma'}\rangle \right]  \nonumber \\ & &+r\lambda=0. \label{mean2}
\end{eqnarray}
We first illustrate possible phases by setting $t'=0$. The energy spectra are found as $E_{\mathbf{k}}=-\mu_{\mathbf{k}} \pm \sqrt{m_{\mathbf{k}}^2+r^2(v_0 \pm 2\lambda_{so}\sqrt{sin^2\mathbf{k}})^2}$.  Clearly, the energy gap is given by $2 \sqrt{m_{\mathbf{k}}^2+r^2(v_0 - 2\lambda_{so}\sqrt{sin^2\mathbf{k}})^2}$. Hence setting $m_{\mathbf{k}}=0$ and $v_0 =2\lambda_{so} \sqrt{sin^2\mathbf{k}}$ determines all gapless phases. More explicitly, gapless phases and corresponding gapless momenta $\mathbf{k}_0$ are determined by $\varepsilon_{\mathbf{k}_0} =(\varepsilon_d+\lambda)/(1+\eta r^2) \equiv\varepsilon_{\lambda}, \label{k0} $ and  $\cos^2\mathbf{k}_0 \equiv \cos^2k_{x0} + \cos^2k_{y0} + \cos^2k_{z0} = 3-(v_0/2\lambda_{so})^2$.
It is clear that the effective parameter that tunes the Kondo lattice through different phases is $\varepsilon_{\lambda}$.  Solutions of $\mathbf{k}_0$ generally form surfaces. As illustrated in Fig.~\ref{fig1}(a), there is a large parameter space that supports gapless phases. In addition to gapless regimes, there are phases that have gaps in electronic structures and are characterized by topological indices $(\nu_0;\nu_1, \nu_2,\nu_3)$\cite{index}. Clearly, going from one gapped phase to another gapped phase with different topological indices, the Kondo lattice has to go through gapless phases.
In the special case when  $v_0=0$, gapless momenta satisfy $cos^2\mathbf{k}_0=3$ so that
$\mathbf{k}_0$ are isolated points at time-reversal invariant momenta: $\Gamma$,$X$, $M$, and $R$. These gapless phases are nodal phases with the transition occurring at $\varepsilon_{\lambda}/t=-6,-2,2,6$ as illustrated Fig.~\ref{fig1}(a).  Since there are 4 degenerate zero-energy states at $\mathbf{k}_0$, these phases are Dirac semi-metallic phases.
In general, $t'$ is non-vanishing and the topological phases are identified in the same way, shown in Fig.~\ref{fig1}(b) for $v_0=0$.  For real materials, $t'/t \sim -0.2$, the STI phase with index $(1;000)$ shrinks, while the $(1;111)$ phase gets enlarged. 

\begin{figure}
\includegraphics[height=1.4in,width=3.5in]{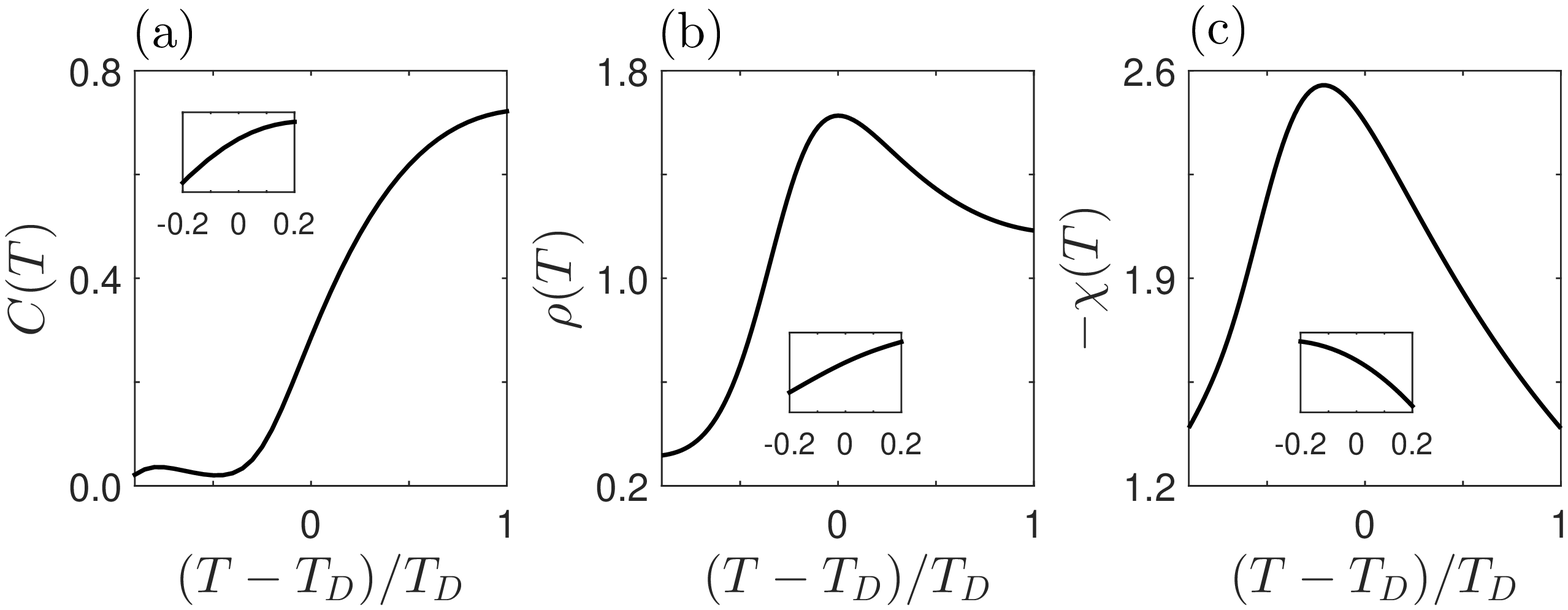}
\caption{Temperature dependence of (a) heat capacity in unit of $k_B$ (b) resistivity in unit of $ 0.1 \pi^2 \hbar^3 v_F/n_D e^2 \epsilon_c$ (c) magnetic susceptibility in unit of $n_D e^2 v_F /3 \pi^2 c^2 \hbar$. Inset figures show the corresponding scaling functions.
Here $C(T)$, $\rho (T)$ and $\chi (T)$ exhibit scaling laws near $T_D$ with exponents $a$ being $3$, $-1$ and $0$ respectively.}\label{thermal}
\end{figure}
%Note that for the special point $t'/t=-0.5$, there is no WTI phase. The transition from $(1;111)$ to %$(1;111)$ goes through a Dirac semi-metallic phase with occurrence of even number of Dirac points.
{\it Fermionic finite-temperature critical point.--}  The electronic structures depend on temperature through $r$ and $\lambda$. As $T$ increases, $r$ decreases and the coupling between $c$ and $d$ electrons decreases. Eventually $c$ and $d$ decouples at $T_K$. Here by taking $U \rightarrow \infty$ and $v_0=0$ in Eq.(\ref{T_K}), $T_K$ is estimated as $T_K = 4t \exp [-9\eta t^2/(2\lambda^2_{so})]$, where we made use of $D_0 \sim 4t$, $ \rho =1/(6t)$, and $\rho_d |\mu-\epsilon_d| \sim 1$ with $\rho_d=1/(6\eta t)$ being density of states of $d$ electrons.

Below $T_K$, the energy spectrum is found by solving Eqs.(\ref{mean1}) and (\ref{mean2}). As indicated in Fig.~\ref{fig1}(a), when $v_0=0$, the system goes through Dirac semi-metallic phases in which the energy spectra reduce to $E_\mathbf{k}^\pm=-\mu_\mathbf{k}\pm \sqrt{m^2_\mathbf{k}+4r^2\lambda^2_{so}sin^2\mathbf{k}}$.
By requiring $m_\mathbf{K}=0$ at time reversal momentum $\mathbf{K}$ in together with Eqs.(\ref{mean1}) and (\ref{mean2}), the resulting system possess Dirac points at some temperatures denoted by $T^*_D$. Here the chemical potential may not be at the Dirac point.  If one further requires that the effective chemical $\mu_\mathbf{K}$ vanishes, the system goes through a semi-metallic phase with nodal points being at $K$ points. The corresponding temperature is denoted by $T_D$, which occurs only at properly tuned chemical potential. Fig.~\ref{TD}(a) shows that indeed the system goes through a  finite-temperature semi-metallic phase at $T=T_D$ at three $X$ points by using proper chemical potential.  Clearly, as shown in the inset, the mechanism for the emergent semi-metallic phase is the occurrence of a band-inversion at $X$ points.

In Fig.~\ref{TD}(b), we show a typical dependence of $T^*_D$ on the electron density $n$ and $\lambda_{so}$. $T^*_D$ generally forms a surface so that Dirac points exist in the electronic structures for a large parameter regime. The red line that cuts through different $n$'s shows the temperature $T_D$. It is clear that finite-temperature Dirac points connect to the zero-temperature QCP smoothly. Hence the line formed by the temperature $T_D$ extends QCP and is a critical line that separates the hole Fermi liquid from the electron Fermi liquid.

The effective Hamiltonian near a Dirac point at $\mathbf{K}$ that occurs at $T^*_D$ can be generally expressed as
\begin{equation}
H_{\mathbf{K}} =\left(
\begin{array}{cc}
\alpha (T-T^*_D) I& \hbar v_F\mathbf{\sigma}\cdot\mathbf{q} \\
 \hbar v_F \mathbf{\sigma}\cdot\mathbf{q}&- \alpha (T-T^*_D) I 
\end{array}
\right)  -\bar{\mu} I ,  \label{effective}
\end{equation}
where $\bar{\mu}$ is the effective chemical potential and $\mathbf{q}=\mathbf{k}-\mathbf{K}$ is the deviation of the momentum from $\mathbf{K}$. $H_{\mathbf{K}}$ is valid within a cutoff with $q<q_c$. When number of electrons are properly tuned so that $\bar{\mu}=0$, $T^*_D$ reduces to $T_D$. At $T_D$, the Kondo lattice is at a finite-temperature critical point. Near $T_D$, any physical response $Q$ exhibits the scaling behavior as
\begin{equation}
Q(T, T_D, n_i, u) = T^{a}  \Phi (\frac{m}{k_B T}, \frac{\bar{\mu}}{k_BT},\frac{\hbar v_F q_c}{k_BT}, \frac{\hbar^3 v^3_F}{n_{im} u^2 k_B T}). \label{scaling}
\end{equation}
%Here $\omega$ is the frequency of the response, 
Here $m= |\alpha (T-T_D)|$, $a$ is the exponent that characterizes $Q$, and $\Phi$ is the universal scaling function.  $n_{im} u^2$ characterizes the disorder strength with $n_{im}$ being the density of impurities and $u$ being the potential strength due to impurities. At $T=T_D$, $m=0$ and Eq.(\ref{scaling}) reduces to the same scaling forms near a zero-temperature QCP. The critical region is thus extends to finite temperatures.

To explore the Fermonic criticality, we examine transport and thermodynamical measurements. For this purpose, it is necessary to include temperature effects due to the quasi-particle lifetime $\tau$. Following \cite{PALee}, we obtain the inverse quasi-particle lifetime as $1/\tau = \left( \frac{rV_K}{\epsilon_d +\lambda - \mu} \right)^2 \frac{(\hbar \omega)^2 + \pi^2 (k_BT)^2}{2(\epsilon_d +\lambda - \mu)}  $ with $\hbar \omega$ being the energy of the quasi-particle\cite{sup}.  By including the self energy of holons in the free energy, one obtains the heat capacity\cite{sup}. The contribution to $C(T)$ due to Dirac points obeys the scaling law with exponent $a=3$\cite{sup}. However, due to the fermionic nature, the contribution to $C(T)$ mainly comes from states near the Fermi energy. Hence, unlike the classical critical point, $C(T)$ is smooth crossing $T=T_D$ as indicated in Fig.~\ref{thermal}(a). On the other hand, the electric transport is determined by available states near the Fermi energy and is expected to be suppressed. By using the Kubo formula, the resistivity $\rho(T)$ is computed in the self-consistent Born approximation\cite{Ando}. The critical exponent is found to be $a=-1$\cite{sup}.
Fig.~\ref{thermal}(b) shows how $\rho(T)$ changes as the system goes through the Dirac semi-metallic phase. A peak is exhibited due to the decreasing density of states. However, due to finite temperature excitations, $\rho$ is not infinite at $T_D$. Similarly,  the magnetic susceptibility $\chi(T)$ also exhibits the Dirac semi-metallic phase. The energy of a Landau level in a magnetic field $B\hat{z}$ is given by $\epsilon^{ss'}_{n,q_z}=-\bar{\mu}+s\sqrt{(E_{q_z}+s'\mu_BB)^2+n \hbar^2\omega_B^2}$ with $s, s'=\pm1$, $q_z$ being the wave-vector along $z$ axis, $\mu_B$ being the Bohr magneton, $E_{q_z} =\sqrt{m^2+\hbar^2 v^2_F q^2_z}$, and  $\hbar\omega_B=\sqrt{2eB v_F^2 / c\hbar}$. $\chi(T)$ is then found by computing the grand potential $\Omega$ and $\chi (T)=-(\frac{\partial^2 \Omega}{\partial B^2})|_{B\rightarrow0}$\cite{sup}.  Fig.~\ref{thermal}(c) shows the computed $\chi(T)$ with scaling exponent being $a=0$. The diamagnetic response is enhanced due to the $n=0$ Landau level that resides at the Fermi energy\cite{Ando}. For 3D Dirac semimetals, the enhancement gets broadened as the $n=0$ Landau level becomes a 1D band along the field direction\cite{sup}.

Finally, we analyze effects on critical scalings due to the off-site Coulomb interaction $
H_C= \frac{e^2}{2 \epsilon} \sum_{i,j} \frac{n(\mathbf{r}_i)n(\mathbf{r}_j)}{|\mathbf{r}_i - \mathbf{r}_j |}$,
where $n(\mathbf{r}_i)$ is the electron density at $i$ site. On the $T^*_D$ surface, the effective Hamiltonian is $H_{\mathbf{K}}+H_C$. As shown in Ref. [\onlinecite{Sheehy}] for graphene, there is a regime dictated by the Dirac point, known as Dirac liquids, in which the Coulomb interaction induces logarithmic corrections in response functions. The dimensionless parameter for the correction is  $\bar{\lambda} = e^2/(\epsilon v_F \hbar)$. Following  Ref. [\onlinecite{Sheehy}], a RG analysis is performed by integrating out modes in $q_c/b<q<q_c$ with $\bar{\lambda (b)}$ and the temperature $T(b)$ on $T^*_D$ surface obeying: $
\frac{d \bar{\lambda}(b)}{d \ln (b)} = - \bar{c} \bar{\lambda} (b)^2$ $\frac{d T(b)}{d \ln b} = T(b) (1- \bar{c} \bar{\lambda} (b) )$, where $\bar{c}=2/3\pi $. Hence $H_C$ is marginal and its effects lie in the regime bounded by the crossover scale $b^*$. Setting $n(b^*)=n_0$ and $T(b^*)=T_0 =\hbar v_F q_c$ with $n_0$ and $T_0$ being electron density and temperature scale in high temperature region\cite{Sheehy}, the crossover temperature is found as
$
T^*(n) = T_0 |n-2|^{1/3} \left( 1+ \bar{c} \frac{\bar{\lambda}}{3} \ln \frac{n_0}{|n-2|} \right).
$
Here $T_0$ weakly depends on $T^*_D$ and hence $T^*$ is roughly a border line. In Fig.~\ref{TD}(b), $T^*$ is indicated by blue dash lines. Inside $T^*$ in the Dirac liquid regime, responses of electrons get corrections by factors of $\Delta(T)= [1+\bar{c} \bar{\lambda} \ln (T_0/T)]$. Replacing $v_F$ by $\Delta(T) v_F$, we find that the diamagnetic susceptibility $\chi$ gets a further enhancement by the factor $\Delta(T)$ in the Dirac liquid regime, while the conductivity and the heat capacity get suppressed by factors of $\Delta(T)$ and $\Delta^3(T)$ respectively.

{\it Discussion and conclusion.--} The fermionic finite-temperature critical point also occurs at 2D\cite{sup}. Experimentally, the critical point can be more easily realized in 2D Kondo lattices, which are formed by introducing adatoms on two dimension materials such as graphene\cite{graphene_kondo}. These critical points are protected as they result from transitions between two topological phases and transitions must go through gapless phases. Therefore, one expects that Dirac semi-metallic phases survive even if fluctuations that are beyond the mean-field theory are included. In real materials, $\lambda_{so}$ and $n$ are fixed so that the system is generally not at QCP. However, our results show that quite generally, by increasing temperature and tuning the electron number, the Kondo lattice will pass through $T_D$. Hence we expect results on measurements shown in Fig.~\ref{thermal} are applicable to materials such as SmB$_6$. In particular, a broad peak in resistivity measurement similar to Fig.~\ref{thermal}(b) was observed in experiments\cite{Fisk}, indicating that the Dirac semi-metallic phase may have been already observed.

%Hence it shows that quite generally, it is easier to reach finite-temperature Dirac semi-metallic phase in the Kondo lattice. 
%To realize the Dirac semi-metallic phases in real materials, we first note that in general, for a given %$\lambda_{SO}$ and $n$, the system may not be right at the quantum critical point in zero-temperature.  %Nonetheless, Fig.~\ref{TD}(c) shows that finite-temperature Dirac points still emerge as one increases %temperature. For instance, for $\lambda_{SO} > \lambda^c_{SO}$ and $n=2$, the system is in the STI %phase at $T=0$. At finite temperatures, Dirac points emerge around $T^*_D/|t| \sim 10^{-3}$. While %these Dirac points may not be right at the Fermi energy, by tuning number of electrons either through %gating or doping, 
In conclusion, we demonstrate that instead of being always a gapped topological insulator below the Kondo temperature shown in Ref.\cite{Coleman}, the Kondo lattice can become gapless by going through a finite-temperature topological transition from a STI phase to a WTI phase. At the transition, the Kondo lattice is a Dirac semimetal, which exhibits finite-temperature relativistic symmetry with nontrivial thermal responses. Our work opens a new pathway to access the Dirac semi-metallic phase and explore the fermionic critical point in the same system.

We acknowledge support 
% Profs. Ming-Che Chang and Sungkit Yip for useful discussions.
by  Ministry of
Science and Technology (MoST), Taiwan.  
%We also acknowledge support by Academia Sinica Research Program on Nanoscience and Nanotechnology, Taiwan.

\end{document}